\documentclass[aps,prc,preprint,showpacs]{revtex4-2}
\usepackage{graphicx}
\usepackage{amsmath}
\usepackage{color}
\begin{document}
\newcommand {\nc} {\newcommand}
\nc {\beq} {\begin{eqnarray}}
\nc {\eol} {\nonumber \\}
\nc {\eeq} {\end{eqnarray}}
\nc {\eeqn} [1] {\label{#1} \end{eqnarray}}
\nc {\eoln} [1] {\label{#1} \\}
\nc {\ve} [1] {\mbox{\boldmath $#1$}}
\nc {\rref} [1] {(\ref{#1})}
\nc {\Eq} [1] {Eq.~(\ref{#1})}
\nc {\re} [1] {Ref.~\cite{#1}}
\nc {\la} {\langle}
\nc {\ra} {\rangle}
\nc {\vJ} {\mbox{$\ve{J}$}}
\nc {\vl} {\mbox{$\ve{l}$}}
\nc {\vL} {\mbox{$\ve{L}$}}
\nc {\vp} {\mbox{$\ve{p}$}}
\nc {\vP} {\mbox{$\ve{P}$}}
\nc {\vq} {\mbox{$\ve{r}$}}
\nc {\vR} {\mbox{$\ve{R}$}}
\nc {\vs} {\mbox{$\ve{s}$}}
\nc {\vS} {\mbox{$\ve{S}$}}
\nc {\vt} {\mbox{$\ve{t}$}}
\nc {\vT} {\mbox{$\ve{T}$}}
\nc {\cL} {\mathcal{L}}
\nc {\cM} {\mathcal{M}}
\nc {\dem} {\mbox{$\frac{1}{2}$}}
\nc {\ut} {\mbox{$\frac{1}{3}$}}
\nc {\qt} {\mbox{$\frac{4}{3}$}}
\nc {\uqa} {\mbox{$\frac{1}{4}$}}
\nc {\arrow} [2] {\mbox{$\mathop{\rightarrow}\limits_{#1 \rightarrow#2}$}}

\author {Ergash M. Tursunov}
\email[]{tursune@inp.uz}
\affiliation {Institute of Nuclear Physics, Uzbekistan Academy of Sciences, \\
100214, Ulugbek, Tashkent, Uzbekistan}
\author{Daniel Baye}
\email{daniel.baye@ulb.be}
\affiliation{Physique Quantique, and Physique Nucl\'eaire Th\'eorique et Physique Math\'ematique, \\
C.P. 229, Universit\'e libre de Bruxelles (ULB), B-1050 Brussels Belgium.}

\title{Analysis of $M1$ capture in the $\alpha(d,\gamma)^6$Li reaction}
\date{\today}

\begin{abstract}
An effective operator is exactly equivalent to the long-wavelength form of the $M1$ operator in transition matrix elements.
It allows us to analytically and numerically analyze the $M1$ contribution to the $\alpha(d,\gamma)^6$Li reaction.
Isoscalar $M1$ transitions from an initial $S$ wave are shown to be forbidden
in radiative capture reactions when distortion is neglected in the initial state.
A calculation in a three-body model with proton, neutron, and a structureless $\alpha$ interacting
through effective forces leads to a negligible $M1$ $S$-factor at small energies.
The dominant $M1$ contribution comes from transitions from an initial $S$ wave to isospin 1 components of the $^6$Li ground state.
It is suggested that using this effective $M1$ operator in other models should clarify
the origin of large discrepancies between $M1$ $S$-factors appearing in the literature.
\end{abstract}

\maketitle

\section{Introduction}
The interest for the $\alpha(d,\gamma)^6$Li radiative capture reaction was first related to the so-called lithium problem in astrophysics.
A discrepancy exists between the observed and calculated $^6$Li abundances in the universe.
Most probably, this problem is not due to inaccuracies in the knowledge of the $\alpha(d,\gamma)^6$Li reaction rate
but it has motivated a number of experiments and theoretical papers.
The LUNA measurements \cite{ATM14,TAA17} at 80, 93, 120 and 133 keV, {\it i.e.}, within the Gamow peak of astrophysical interest,
have provided strict constraints for testing the validity of theoretical radiative capture calculations
and restricted the probability of a nuclear physics solution of the lithium problem.
The origin of the discrepancy should rather be found in an astrophysical context \cite{ALN06,LMA13,FO22}.

The study of $\alpha(d,\gamma)^6$Li radiative capture remains nevertheless very interesting by itself
because it requires solving a number of delicate problems crucial for reactions involving $N = Z$ nuclei.
The conclusions from its analysis will be essential for future accurate microscopic studies
of the important $^{12}$C($\alpha,\gamma)^{16}$O reaction.

In reactions  involving $N = Z$ nuclei, three multipolarities, namely $E1$, $E2$, and $M1$, enter in competition.
Indeed, due to an approximate selection rule, the usually dominant $E1$ transitions are forbidden
between isospin-zero states by an isospin selection rule.
At the long-wavelength approximation (LWA), the isoscalar part of the $E1$ operator vanishes
and transitions only take place via its isovector part.
Matrix elements of isovector operators vanish between $T = 0$ states.
However, except for the deuteron, realistic wave functions of $N = Z$ nuclei are not pure $T = 0$ states
and isovector $E1$ transitions are thus not exactly forbidden.
Moreover, isoscalar $E1$ transitions remain possible beyond the LWA \cite{Ba12,Ba23}.
As a result, the $E1$ strength may keep an order of magnitude similar
to the strengths of the usually much weaker $E2$ and $M1$ transitions.

Each multipolarity appearing in the $\alpha(d,\gamma)^6$Li capture leads to theoretical difficulties.
For $E1$ capture, the beyond LWA isoscalar part of the  matrix element
and its isovector part requiring isospin mixing add coherently.
The isovector partial matrix element requires taking account of small $T = 1$ components in the final $^6$Li and initial scattering wave functions.
The $E2$ capture can take place up to very large distances and imposes a good reproduction of the asymptotic normalization constant (ANC) of the $^6$Li ground state.
A correct treatment of $M1$ capture requires calculations respecting
the perfect orthogonality between the initial and final wave functions \cite{NWS01}.

The relative importance of these electromagnetic components is still an open question
as existing models \cite{NWS01,BT18,TTK18,TTK20,So22,HHK22} have given various results.
In \re{NWS01}, these components are evaluated in a model involving a realistic six-body final wave function
and an initial wave function involving realistic clusters and a phenomenological relative motion.
As isospin mixing is neglected, the forbidden $E1$ transitions are treated phenomenologically.
The $M1$ contribution is found negligible with respect to $E1$ and $E2$.
Only $E1$ and $E2$ captures are studied with a $p+n+\alpha$ non-microscopic model in \re{BT18}.
The $E2$ component is improved in \re{TTK18} by adjusting the ANC.
The $E1$ results involve a large isovector contribution and a small isoscalar part.
However, the $T=1$ $^6$Li component in the model is very sensitive to the way
forbidden states of the effective $\alpha N$ potential are eliminated \cite{TTK20} and may be overestimated.
In \re{BT18}, forbidden states in the $^6$Li final wave function are removed with an orthogonalizing pseudopotential \cite{KP78}.
When elimination is performed with a supersymmetric  transformation \cite{Ba87,DDB03} in place of an orthogonalizing pseudopotential,
the size of the isospin content in $^6$Li is reduced and the $E1$ contribution is an order of magnitude smaller \cite{TTK20}.
In \re{So22}, the three multipolarities are computed within a two-center microscopic cluster model
with several parametrizations of the nucleon-nucleon ($NN$) interaction.
The wave function is an antisymmetrized product of harmonic oscillator ground state functions around each center.
The wave functions are purely isopin 0 and the isovector $E1$ contribution is therefore absent.
Both $E1$ and $M1$ contributions are essentially negligible with respect to $E2$.
In \re{HHK22}, an {\it ab initio} calculation is performed within the no-core shell model with continuum
with $NN$ and three-nucleon interactions.
The $E1$ component is smaller than in all previously published calculations
but the authors find that $M1$ capture is an important part of the total $S$-factor at low energies.
The $M1$ $S$-factor even increases at very small energies.
This large $M1$ component is in contradiction with other evaluations which suggest that it is negligible \cite{NWS01,So22}.
This controversy deserves a clarification.

The aim of the present paper is to analyze the structure of the $M1$ component of $\alpha(d,\gamma)^6$Li radiative capture.
To this end, we make use of an effective $M1$ operator strictly equivalent to the LWA form of this operator
when the initial and final states are orthogonal \cite{Ba23}.
The structure of matrix elements of this equivalent operator clarifies properties of the various types of $M1$ transitions.
The numerical importance of the different multipolarities are then evaluated in a three-body model where isospin mixing is simulated
\cite{BT18,TTK20}.

In Sec.~\ref{sec:M1}, an effective operator is shown to be equivalent to the LWA form of the $M1$ operator in radiative capture reactions.
It is used to determine the main allowed $M1$ transitions at small energies for initial scattering states without distortion.
In Sec.~\ref{sec:tbm}, these properties are tested and illustrated in a $p + n + \alpha$ three-body model with effective forces.
Consequences are discussed in Sec.~\ref{sec:dis}.
Sec.~\ref{sec:conc} is devoted to a conclusion.

Except in \Eq{2.1}, we use $\hbar = c = 1$.
\section{Effective $M1$ transition operator}
\label{sec:M1}
\subsection{Microscopic $M1$ operator}
\label{sec:mo}
At the LWA, the magnetic dipole operator for pointlike nucleons is given in translation-invariant form by
(see Eq.~(50) of \re{Ba23} where $\hbar = 1$)
\beq
\cM_\mu^{M1} = \frac{\mu_{\rm N}}{\hbar c} \sqrt{\frac{3}{4\pi}} \sum_{j=1}^A \left( g_{lj} L'_{j\mu} + g_{sj} S_{j\mu} \right),
\eeqn{2.1}
where $\mu_{\rm N} = e\hbar/2m_p$ is the nuclear magneton involving the proton mass $m_p$.
The orbital angular momentum of nucleon $j$ with respect to the center of mass of the $A$ nucleons and its spin
are denoted as $\vL'_j$ and $\vS_j$, respectively.
In terms of the third component $t_{j3}$ of the isospin $\vt_j$ of nucleon $j$, the coefficients read
$g_{lj} = \dem - t_{j3}$ and $g_{sj} = g_p (\dem - t_{j3}) + g_n (\dem + t_{j3})$
as a function of the proton $g_p$ and neutron $g_n$ gyromagnetic factors.
The $M1$ operator can be rewritten as a sum of isoscalar and isovector terms
\beq
\cM_\mu^{M1} = \mu_{\rm N} \sqrt{\frac{3}{4\pi}}
\left\{ \frac{1}{2} \left[ \cL_\mu + (g_p + g_n) S_\mu \right] - \sum_{j=1}^A t_{j3} \left[ L'_{j\mu} + (g_p - g_n) S_{j\mu} \right] \right\},
\eeqn{2.2}
where $\ve{\cL} = \sum_{j=1}^A \vL'_j$ is the total orbital momentum operator of the system
and $\vS = \sum_{j=1}^A \vS_j$ is its total spin.
The total angular momentum is $\vJ = \ve{\cL} + \vS$.
The total isospin operator is $\vT = \sum_{j=1}^A \vt_j$.

Operator \rref{2.1} is usually employed in radiative capture calculations \cite{NWS01,So22,HHK22}.
Here we replace it by an operator which is exactly equivalent in those calculations
but allows a deeper analysis of the structure of $M1$ transitions.
To this end, we take advantage of the facts that initial and final states of a transition are eigenstates of the total angular momentum of the system
and are therefore orthogonal in radiative capture reactions because of different angular momenta or parities, or otherwise of different energies.
Indeed, the matrix elements
\beq
\la \Psi_f^{JM\pi} | J_\mu | \Psi_i^{J'M'\pi'} \ra = 0
\eeqn{2.3}
of components $J_\mu$ of the total angular momentum between the initial and final wave functions $\Psi_i^{J'M'\pi'}$ and $\Psi_f^{JM\pi}$
always vanish because of this orthogonality.
Hence, the operator
\beq
\widetilde{\cM}_\mu^{M1} = \cM_\mu^{M1} - \mu_{\rm N} \sqrt{\frac{3}{4\pi}} \frac{1}{2} J_\mu
\eeqn{2.4}
has the same radiative capture matrix elements as operator \rref{2.1}.
This operator contains isoscalar and isovector parts.
It has the explicit form (51) of \re{Ba23},
\beq
\widetilde{\cM}_\mu^{M1} = \mu_{\rm N} \sqrt{\frac{3}{4\pi}}
\left\lbrace  \frac{1}{2} (g_p + g_n -1) S_\mu - \sum_{j=1}^A t_{j3} \left[ L'_{j\mu} + (g_p - g_n) S_{j\mu} \right] \right\rbrace
\eeqn{2.5}
with $g_p + g_n - 1 \approx 0.760$ and $g_p - g_n \approx 9.412$.
Only the total spin operator of the nucleons appears in the isoscalar first term.
The isovector part is unchanged with respect to \Eq{2.2}.
\subsection{Transition matrix elements}
\label{sec:tme}
Let us turn to $\alpha + d$ capture.
The $^6$Li states can be decomposed in $T = 0$ and $T = 1$ isospin components as
\beq
\Psi_f^{JM\pi} = \Psi_f^{JM\pi;0} + \Psi_f^{JM\pi;1}.
\eeqn{2.6}
Here possible small higher isospin components are neglected.
In the entrance channel, the $\alpha + d$ scattering state has a total intrinsic spin $S = 1$.
Its total angular momentum $J$ results from the coupling of this spin $S = 1$ with the relative orbital momentum $L$ between $\alpha$ and deuteron and its parity is $(-1)^L$.
Allowed $M1$ transitions from $\alpha + d$ initial scattering states to the final $^6$Li$(1^+)$ ground state are thus possible only from the $J^\pi = 1^+$ and $2^+$ initial partial waves.

For $J^\pi = 1^+$, we assume that the initial wave function is a sum of $L = 0$ and $L = 2$ terms,
\beq
\Psi_i^{1M+} = \Psi_{i0}^{1M+} + \Psi_{i2}^{1M+}.
\eeqn{2.7}
Distortion at small relative distances is thus neglected.
This initial state has a total spin $S = 1$ and total isospin $T = 0$.
Property
\beq
S_z \Psi_i^{1M+} = M \Psi_{i0}^{1M+} + S_z \Psi_{i2}^{1M+}
\eeqn{2.8}
leads with \Eq{2.6} to the transition matrix element
\beq
\la \Psi_f^{1M+} | \cM_z^{M1} | \Psi_i^{1M+} \ra & = & \la \Psi_f^{1M+} | \widetilde{\cM}_z^{M1} | \Psi_i^{1M+} \ra
\eol
& = & \mu_{\rm N} \sqrt{\frac{3}{4\pi}} \left\{ \frac{1}{2} (g_p + g_n -1) \la \Psi_f^{1M+;0} | S_z - M | \Psi_{i2}^{1M+} \ra \right.
\eol
&& \left. - \la \Psi_f^{1M+;1} | \sum_{j=1}^A t_{j3} \left[ L'_{jz} + (g_p - g_n) S_{jz} \right] | \Psi_i^{1M+} \ra \right\}
\eeqn{2.9}
since $\Psi_f^{1M+;0}$ is orthogonal to $\Psi^{1M+}_{i0} + \Psi^{1M +}_{i2}$ because of the isospin $T = 0$ of the initial state.
One obtains the important result that isoscalar transitions from an initial $S$ wave are forbidden
when distortion is neglected in this wave or, more generally, when $S_z \Psi_{i0}^{1M+} = M \Psi_{i0}^{1M+}$.

Isovector transitions from an initial $S$ partial wave remain possible.
Given that $S$ components become larger than $D$ components at very small energies,
$\Psi^{1M+}_{i0}$ can play a more important role than $\Psi^{1M+}_{i2}$ in the second term of \Eq{2.9}.
At small energies, the value of $M1$ transition matrix elements for $\alpha + d$ radiative capture mainly
arises from the small $D$ component of the initial scattering wave function to the isospin 0 part of the final state
or from its $S$ component of the isospin 1 part.

Equation \rref{2.9} is also valid for $J^\pi = 2^+$ since $\Psi_i^{2M+} \equiv \Psi_{i2}^{2M+}$ is orthogonal to $\Psi_f^{1M+}$.

These properties are now studied in the simplest model with an isospin 1 component in the final state.
\section{Three-body model of $M1$ transitions}
\label{sec:tbm}
\subsection{Three-body $M1$ operator}
\label{sec:tbmo}
In this section, we analyze $M1$ transitions in $\alpha + d$ radiative capture
in a three-body model \cite{DDB03,TDB07,TTK16,BT18,TTK18,TTK20}.
The three particles $p$, $n$, and $\alpha$ are assumed to be pointlike.
The $\alpha$ particle has spin 0 and positive parity.
The relative coordinate between neutron and proton is denoted as $\vq = \vq_n - \vq_p$
and the coordinate of the center of mass of these particles with respect to the $\alpha$ particle is denoted
as $\vR = \vq_\alpha - \dem (\vq_p + \vq_n)$.
The corresponding momenta and orbital momenta are denoted as $\vp$, $\vP$ and $\vl$, $\vL$, respectively.
This model involves $\alpha N$ and $NN$ effective interactions.

For $^6$Li, Eq.~(B1) of \re{DDB03} can be summarized as
\beq
\cM_\mu^{M1} = \sqrt{\frac{3}{4\pi}} \left[ \mu_{\rm N} \left(L'_{p\mu} + \frac{1}{2} L'_{\alpha\mu} \right)
+ 2 \sum_{i=1}^3 \mu_i S_{i\mu} \right],
\eeqn{3.1}
where the proton is particle 1 with $A_1 = Z_1 = 1$ and magnetic moment $\mu_1 = \dem g_p \mu_{\rm N}$,
the neutron is particle 2 with $A_2 = 1$, $Z_2 = 0$, and magnetic moment $\mu_2 = \dem g_n \mu_{\rm N}$,
and the $\alpha$ is particle 3 with $A_3 = 4$, $Z_3 = 2$ and magnetic moment $\mu_3 = 0$.
The coefficients in front of $L'_{i\mu}$ arise from factors $Z_i/A_i$.
A missing factor of 2 in Eq.~(B1) of \re{DDB03} is reintroduced in front of the spin terms.

The $M1$ operator can be written as
\beq
\cM_\mu^{M1} & = & \mu_{\rm N} \sqrt{\frac{3}{4\pi}} \frac{1}{2} \{ [L'_{p\mu} + L'_{n\mu} + L'_{\alpha\mu} + (g_p + g_n) S_\mu]
\eol
&& + [ L'_{p\mu} - L'_{n\mu} + (g_p - g_n) (S_{p\mu} - S_{n\mu})] \},
\eeqn{3.2}
where $\vS = \vS_p + \vS_n$ is considered as the total spin since the $\alpha$ particle has no spin.
The first square bracket is symmetric with respect to the exchange of proton and neutron and the second one is antisymmetric.
They correspond to the isoscalar and isovector parts in \Eq{2.2}, respectively.

The effective three-body $M1$ operator corresponding to \Eq{2.5} reads
\beq
\widetilde{\cM}_\mu^{M1} = \mu_{\rm N} \sqrt{\frac{3}{4\pi}} \frac{1}{2} \{ (g_p + g_n - 1) S_\mu
+ [ L'_{p\mu} - L'_{n\mu} + (g_p - g_n) (S_{p\mu} - S_{n\mu})] \}.
\eeqn{3.3}
As for the isoscalar part in \Eq{2.5}, the symmetric part only involves the total spin.
Comments about allowed transitions at the end of Subsec.~\ref{sec:tme} still hold.
\subsection{Three-body wave functions}
\label{sec:tbwf}
The wave functions are approximate variational solutions of $p + n + \alpha$ Hamiltonians.
An isospin dependence is simulated by the $p + n$ part of the wave functions.
Indeed, $p + n$ wave functions must be antisymmetric in the isospin formalism.
They must obey the selection rule: $l + S + T$ odd.
Components of the three-body wave functions with $l + S$ odd correspond to $T = 0$
and components with $l + S$ even correspond to $T = 1$.

The $^6$Li$(1^+)$ ground state wave function is obtained with a three-body Hamiltonian with effective $NN$ and $\alpha N$ interactions.
This $p + n +\alpha$ wave function is calculated in hyperspherical coordinates \cite{DDB03}.
As explained in the Appendix, it can be expanded as
\beq
\Psi_f^{J_fM_f+} = \sum_\gamma [[Y_{l_f}(\Omega_r) \otimes Y_{L_f}(\Omega_R)]^{\cL_f} \otimes \chi_{S_f}]^{J_fM_f} \psi^{J_f}_{f\gamma}(r,R),
\eeqn{3.4}
where $J_f$ is equal to 1, $\gamma$ represents $(l_fL_f)\cL_fS_f$, and $\chi_{S_f}$ is the total spinor. Positive parity imposes $l_f + L_f$ even.
Some terms are symmetric with respect to the exchange of proton and neutron and correspond to isospin 0
while other terms are antisymmetric and correspond to isospin 1.
For example, $S_f = 1$ and $l_f = 1$, or $S_f = 0$ and $l_f = 0$ or 2, simulate isospin $T = 1$.

The initial $p + n + \alpha$ partial scattering waves are described differently.
They are solutions of a separable Hamiltonian in $\vR$ and $\vq$ coordinates involving the same $NN$ interaction as for the final state and an effective $\alpha + d$ interaction (see \re{TTK16,BT18}).
Initial wave functions are thus described as coupled products of the ground state deuteron wave function
with internal orbital momentum $l = 0$ and total spin $S = 1$ and $\alpha + d$ scattering waves.
Distortion of the $p + n$ part is neglected.
These scattering functions involve the deuteron ground state wave function with $l + S = 1$ and thus correspond to isospin $T = 0$.
Only $J^\pi = 1^+$ and $2^+$ initial states are possible and allowed for a transition to a $1^+$ final state.
As positive parity is conserved in the $M1$ transition, they are defined as
\beq
\Psi_i^{JM+} = \sum_{L\ {\rm even}} Y_0^0(\Omega_r) [Y_L(\Omega_R) \otimes \chi_1]^{JM} \psi^J_{iL}(r,R),
\eeqn{3.5}
where
\beq
\psi^J_{iL}(r,R) = (rR)^{-1} u_d(r) g_L^J(R).
\eeqn{3.6}
In \Eq{3.6}, $u_d(r)$ represents the deuteron radial function derived from the effective $NN$ interaction
and $g_L^J(R)$ is a radial function of the $\alpha + d$ relative motion derived from a central potential,
with asymptotic behavior
\beq
g_L^J(R) \arrow{R}{\infty} \cos \delta_L^J F_L(kR) + \sin \delta_L^J G_L(kR),
\eeqn{3.7}
where $k$ is the wavenumber, $\delta_L^J$ is the phase shift, and $F_L$ and $G_L$ are the regular and irregular Coulomb functions.
For $1^+$, the wave function contains two terms, $L = 0$ and 2.
Without tensor force, $g_0^1$ and $g_2^1$ are decoupled.
For $2^+$, only $g_2^2(R)$ appears.

As the initial and final wave functions are not arising from the same Hamiltonian,
the $J = 1$ initial wave function \rref{3.5} must be orthogonalized to the $J_f = 1$ $^6$Li ground-state wave function \rref{3.4}.
Function $\psi^J_{iL}(r,R)$ is thus modified in this case.
See Appendix for details.
\subsection{Three-body $M1$ matrix elements}
\label{sec:tbme}
In coordinates $\vq$ and $\vR$, the $M1$ operator \rref{3.1} is given by Eq.~(B5) of \re{DDB03},
\beq
\cM_\mu^{M1} = \mu_{\rm N} \sqrt{\frac{3}{4\pi}}
\left\{ \frac{Z_{12}A_3^2 + Z_3A_{12}^2}{AA_{12}A_3} L_\mu + \frac{Z_1A_2^2 + Z_2A_1^2}{A_1A_2A_{12}} l_\mu \right.
\eol
\left. + \left( \frac{Z_1}{A_1} - \frac{Z_2}{A_2} \right)
\left[ \frac{A_3}{A} (\vR \times \vp)_\mu + \frac{A_1A_2}{A_{12}^2} (\vq \times \vP)_\mu \right]
+\sum_{i=1}^3 g_i S_{i\mu} \right\}
\eeqn{3.9}
where $Z_{12} = Z_1 + Z_2$, $A_{12} = A_1 + A_2$, and $A = A_{12} + A_3$.
Note that misprints need be corrected in Eq.~(B5):
$\vl_x$ and $\vl_y$ must be exchanged and a factor of 2 is missing in front of the spin term.
In the $p + n +\alpha$ case, \Eq{3.9} becomes
\beq
\cM_\mu^{M1} = \mu_{\rm N} \sqrt{\frac{3}{4\pi}} \frac{1}{2}
\left[ L_\mu + l_\mu + \frac{4}{3} (\vR \times \vp)_\mu + \frac{1}{2} (\vq \times \vP)_\mu
+ 2(g_p S_{p\mu} + g_n S_{n\mu}) \right]
\eeqn{3.10}
since the $\alpha$ particle has no spin.

With \Eq{2.4} where $\vJ = \vL + \vl + \vS$, the effective operator \rref{3.3} is given by
\beq
\widetilde{\cM}_\mu^{M1} & = & \mu_{\rm N} \sqrt{\frac{3}{4\pi}} \frac{1}{2} \left[ (g_p + g_n - 1) S_\mu \right.
\eol
&& \left. + \frac{4}{3} (\vR \times \vp)_\mu + \frac{1}{2} (\vq \times \vP)_\mu + (g_p - g_n) (S_{p\mu} - S_{n\mu}) \right].
\eeqn{3.11}
The first term is symmetrical with respect to the exchange of neutron and proton and is thus isoscalar.
The spin scalar and spin vector parts of the second term are antisymmetrical and thus isovector.

According to Eqs.~\rref{2.9} and \rref{3.11}, the transition matrix element reads
\beq
&& \la \Psi_f^{1M+} | \cM_z^{M1} | \Psi_i^{1M+} \ra
= \mu_{\rm N} \sqrt{\frac{3}{4\pi}} \left\{ \frac{1}{2} (g_p + g_n -1) \la \Psi_f^{1M+;0} | S_z - M | \Psi_{i2}^{1M+} \ra \right.
\eol
& - & \left. \la \Psi_f^{1M+;1} | \frac{4}{3} (\vR \times \vp)_z + \frac{1}{2} (\vq \times \vP)_z
+ (g_p - g_n) (S_{pz} - S_{nz}) | \Psi_i^{1M+} \ra \right\}.
\eeqn{3.12}
An isoscalar $M1$ transition from the $S$ wave is forbidden.
An isovector $M1$ transition from the $S$ wave is allowed at least toward component $S_f = l_f = L_f = 1$ in the final wave function.
\subsection{Results}
\label{sec:res}
The deuteron radial wave function $u_d(r)$ is computed in a two-body model
with the central Minnesota interaction of \re{TLT77} as in Refs.~\cite{TTK16,BT18,TTK18,TTK20}.
The Schr\"odinger equation is solved with the Lagrange-mesh method \cite{Ba15} with $\hbar^2/2m_N = 20.7343$ MeV fm$^2$.
The scattering wave functions of the $\alpha + d$ relative motion are calculated with a deep potential of \re{DD94}
slightly modified in the $S$ wave as $V_{\alpha d}^{L=0}(R) = -92.44 \exp(-0.25 R^2)$ MeV
to reproduce the empirical value $C_{\alpha d} = 2.31$ fm$^{-1/2}$ of the ANC of the $^6$Li ground state \cite{BKS93},
and unchanged in other partial waves.

As in \re{BT18}, a three-body Hamiltonian is solved for the final wave function.
The $^6$Li$(1^+)$ ground-state wave function is obtained within an approximately variational calculation \cite{DDB03}.
The central Minnesota $NN$ potential is also employed as neutron-proton interaction \cite{TLT77}.
For the $\alpha N$ nuclear interaction, the deep potential of \re{VKP95} is employed
slightly renormalized by 1.014 in the $S$ wave to reproduce the experimental binding energy 3.70 MeV of $^6$Li
with respect to the $p + n + \alpha$ threshold.
The Coulomb interaction between $\alpha$ and proton is taken as $2e^2\, \mathrm{erf}(0.83\, R)/R$ \cite{RT70}.
The $\alpha N$ interaction contains a forbidden state which must be eliminated in a three-body model \cite{DDB03}.
This elimination is performed with the method of orthogonalizing pseudopotentials \cite{KP78}
as in Refs.~\cite{VKP95,KPT96,DDB03,TDB07,TTK16,BT18,TTK18}.
The hypermomentum expansion includes terms up to $K_{\rm max} = 24$.
The coupled hyperradial equations are solved with the Lagrange-mesh method \cite{DDB03,Ba15}.
The isotriplet component in the $^6$Li ground state has a squared norm $5.3 \times 10^{-3}$ \cite{TTK18}.

The partial radiative capture cross section due to $M1$ transitions is given by \cite{DB10}
\beq
\sigma^{M1}(E) = \frac{64\pi^2}{9\hbar v k^2} k_{\gamma}^3
\sum_J \sum_L \left| \sum_{T = 0}^1 \la \Psi_f^{1+;T} || \cM^{M1} || \Psi^{J+}_{iL}(E) \ra \right|^2,
\eeqn{4.1}
where $E$ and $v$ are the relative energy and velocity between $\alpha$ and deuteron, $k_\gamma$ is the photon wavenumber,
and $\Psi^{JM+}_{iL}$ is term $L$ of the sum in \Eq{3.5}.
The $T = 0$ and 1 parts of the final wave function are
separately shown for later use.
From this cross section the $M1$ $S$-factor is given by $S^{M1}(E) = E\sigma^{M1}(E) \exp(2\pi\eta)$ where $\eta$ is the Sommerfeld parameter.

We now discuss the effects of the different components in the matrix elements.
Isoscalar transitions from an initial $S$ wave are exactly forbidden by the effective $M1$ operator \rref{3.11} in the present model.
We have checked numerically that they are correctly forbidden by the $M1$ operator \rref{3.10}, as expected.

\begin{figure}[ht]
\setlength{\unitlength}{1 mm}
\begin{picture}(160,90) (0,0)
\put(20,0){\mbox{\scalebox{0.4}{\includegraphics{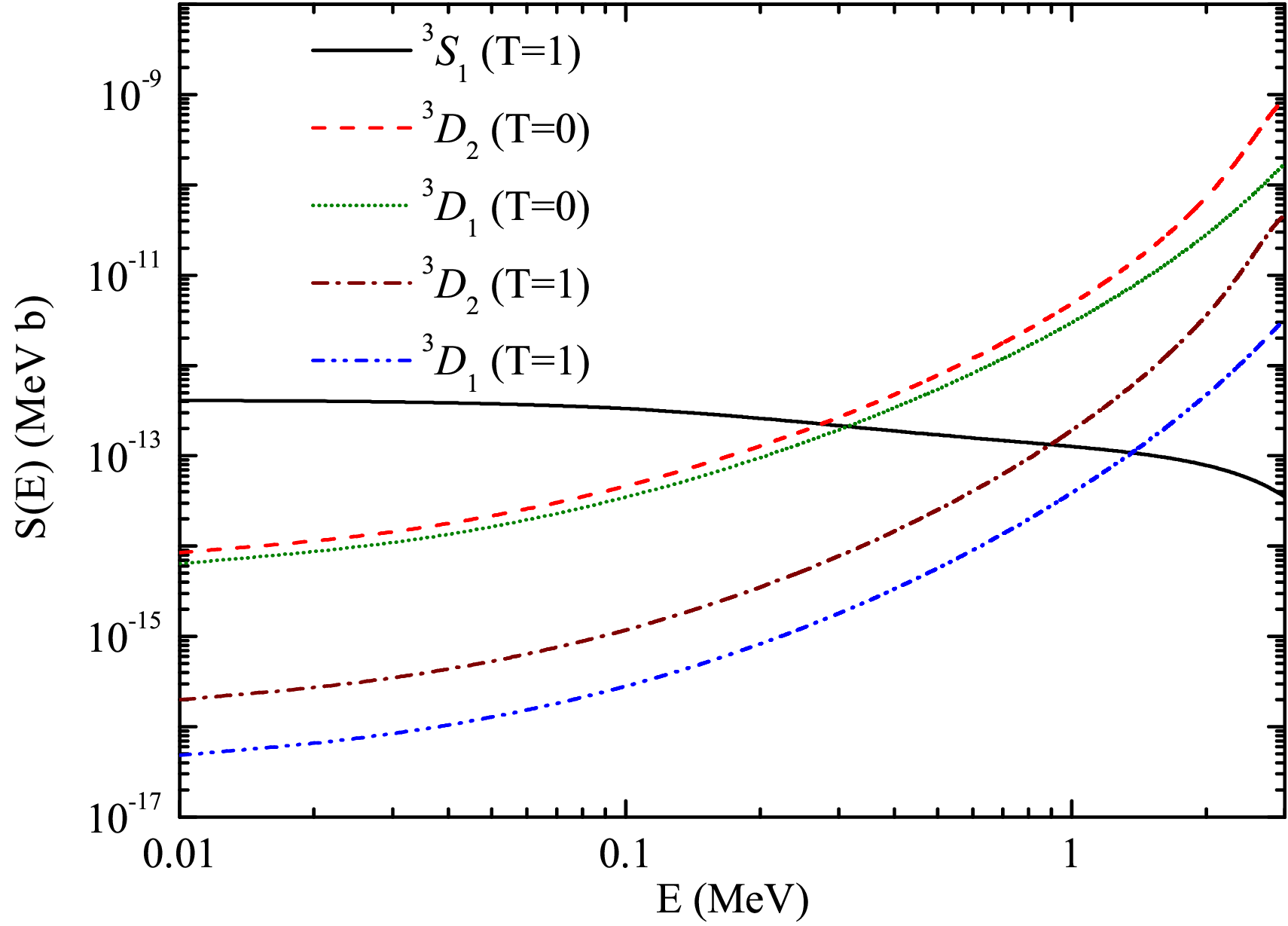}}}}
\end{picture} \\
\caption{Size of $T = 0$ and $T = 1$ isospin components displayed as $M1$ $S$-factors.}
\label{fig:1}
\end{figure}
In \Eq{4.1}, the matrix elements corresponding to $T = 0$ and $T = 1$ in the final wave function,
\textit{i.e.}, simulated by symmetry and antisymmetry with respect to the exchange of proton and neutron in the $M1$ operator, add coherently.
The sizes of these isospin terms are first analyzed by dropping the sum over $T$ in \Eq{4.1}.
The $S$-factors obtained from separate isoscalar and isovector parts of matrix element \rref{3.12} are displayed in Fig.~\ref{fig:1}.
The purely isovector $^3S_1$ $S$-factor is presented as a full curve.
The behavior of this spin scalar transition is typical for an initial $S$ wave:
it is rather flat at small energies and becomes much larger than components from other initial waves at very low energies.
In the same figure are shown the isoscalar $^3D_1$ and $^3D_2$ transitions.
As expected for $D$ transitions, they decrease fast when $E \rightarrow 0$.
The isovector $^3D_1$ and $^3D_2$ curves are still weaker by about two orders of magnitude at the lowest energies.
This smallness is expected from the fact that they reach a small $T = 1$ component.
The $T = 1$ $^3D_J$ \textit{matrix elements} are smaller by about one order of magnitude than the $T = 0$ ones.

\begin{figure}[ht]
\setlength{\unitlength}{1 mm}
\begin{picture}(160,85) (0,0)
\put(20,0){\mbox{\scalebox{0.4}{\includegraphics{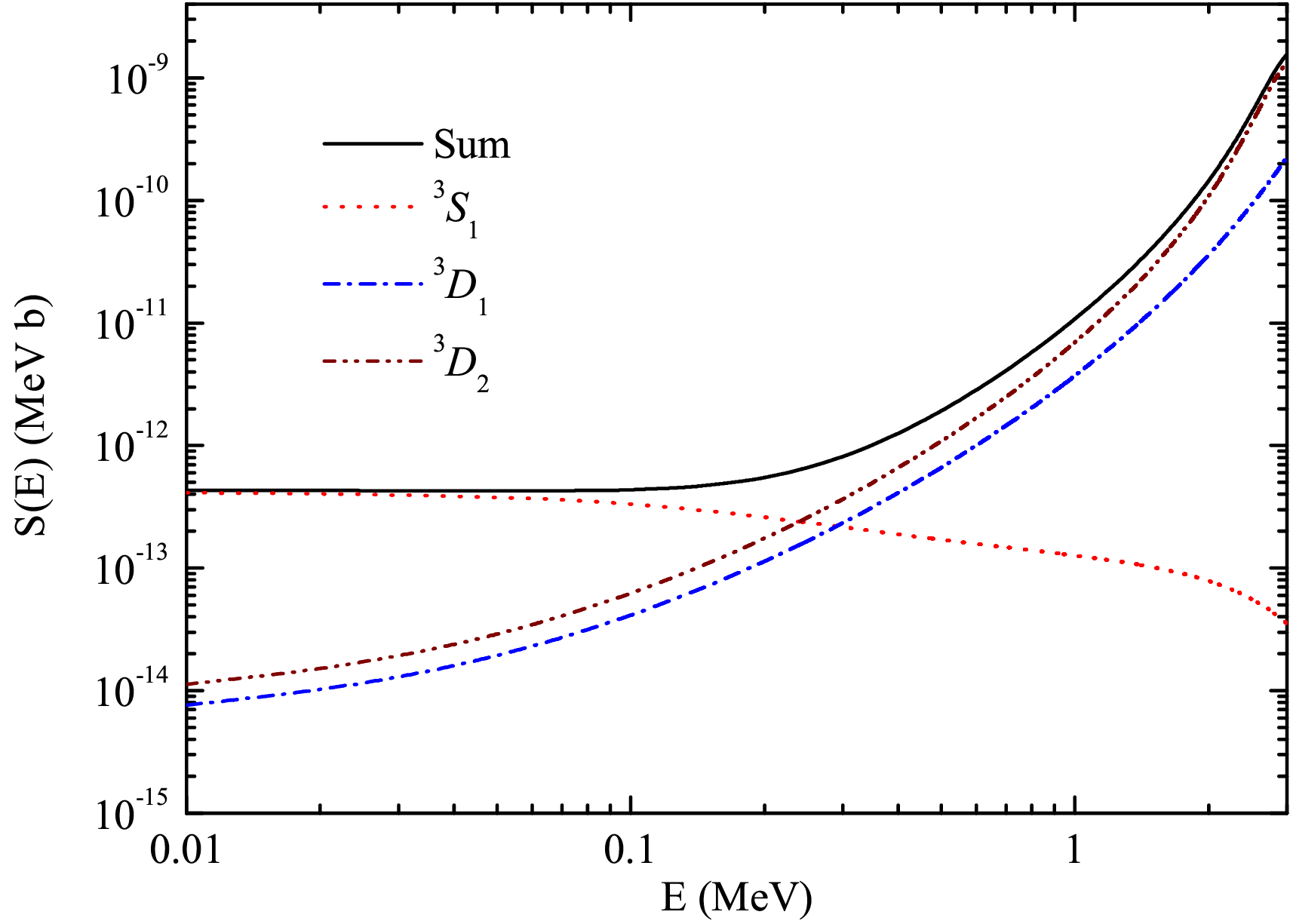}}}}
\end{picture} \\
\caption{Partial $M1$ $S$-factors from the $^3S_1$, $^3D_1$, and $^3D_2$ initial partial waves.}
\label{fig:2}
\end{figure}
Partial $S$-factors with interference effects in the full matrix elements are depicted in Fig.~\ref{fig:2}.
The isovector $^3S_1$ component dominates at small energies.
Isoscalar and isovector $^3D_J$ components add coherently but their interference does not much modify the isoscalar $^3D_J$ values.

\begin{figure}[ht]
\setlength{\unitlength}{1 mm}
\begin{picture}(160,80) (0,0)
\put(20,0){\mbox{\scalebox{0.4}{\includegraphics{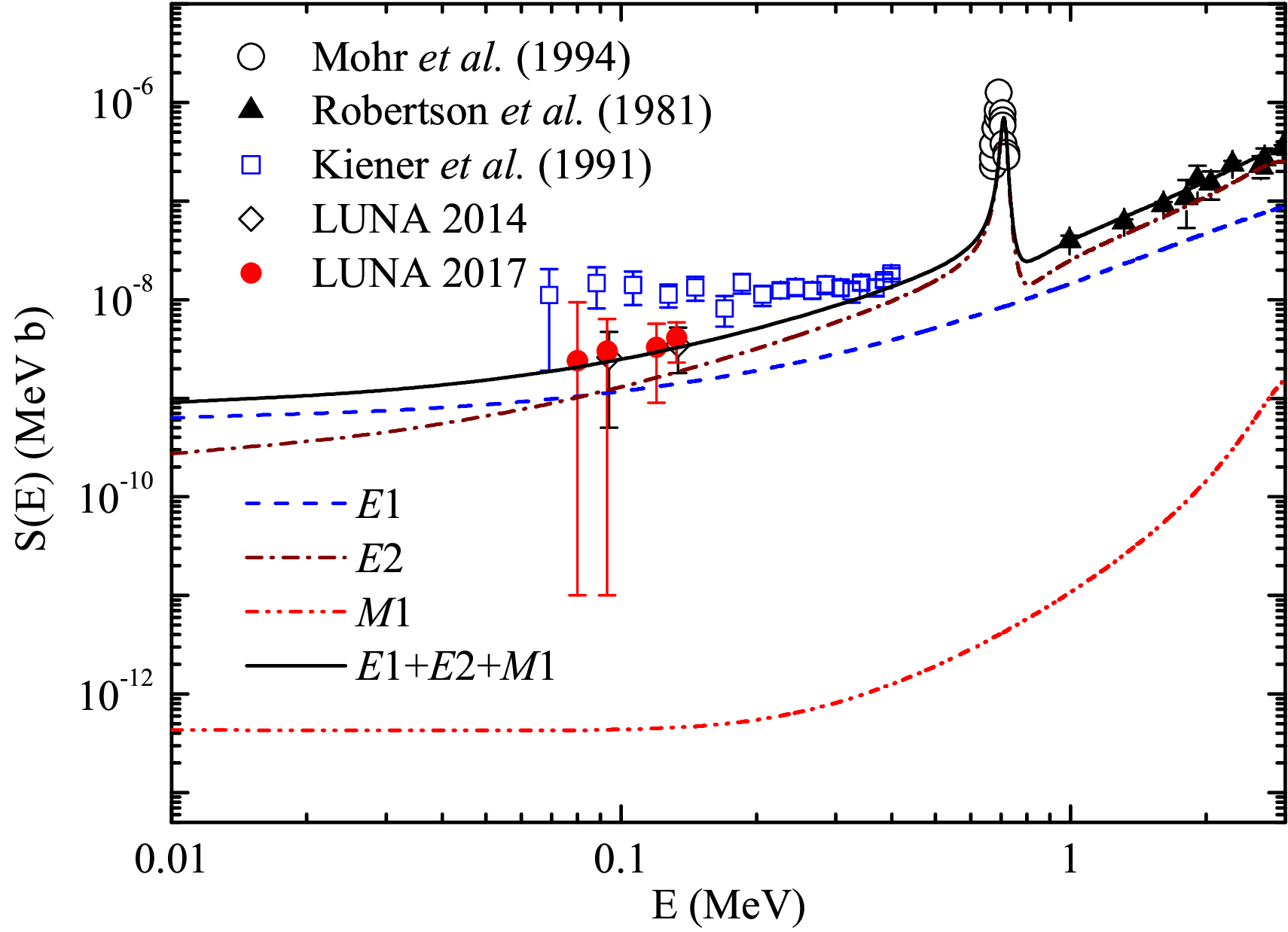}}}}
\end{picture} \\
\caption{$E1$, $E2$, and $M1$ $S$-factors of the $\alpha(d,\gamma)^6$Li reaction and their sum (full line).
The $E1$ and $E2$ contributions are obtained as in Refs.~\cite{BT18} and \cite{TTK18}, respectively,
compared with experimental data of Robertson \textit{et al} \cite{RDW81}, Kiener \textit{et al} \cite{KGR91},
Mohr \textit{et al} \cite{MKW94}, Anders  \textit{et al} (LUNA14) \cite{ATM14}, and Trezzi \textit{et al} (LUNA17) \cite{TAA17}. }
\label{fig:3}
\end{figure}
In Fig.~\ref{fig:3}, the present $M1$ results are compared with $E1$ and $E2$ contributions
obtained as in Refs.~\cite{BT18} and \cite{TTK18}, respectively.
Let us recall here that, while the size of the $E2$ $S$-factor is globally consistent with other references,
the $E1$ $S$-factor is significantly larger.
The important information is that the $M1$ $S$-factor is still orders of magnitude smaller than the dominant $E2$ transitions.
Its contribution to the total $S$-factor is negligible in the present study
in contradiction with the results of \re{HHK22} but in agreement with those of Refs.~\cite{NWS01,So22}.
\section{Discussion}
\label{sec:dis}
In radiative capture reactions, operator $\widetilde{\cM}_\mu^{M1}$ in \Eq{2.5} is equivalent to -- but simpler than --
the magnetic dipole operator $\cM_\mu^{M1}$ for pointlike particles in \Eq{2.1}.
It can also be useful to compute nuclear magnetic dipole moments by adding the matrix elements
of $\widetilde{\cM}_z^{M1}$ and a term proportional to the $z$ component of the total angular momentum as in \Eq{2.4}.

The main information provided by the equivalent effective $M1$ operator is that its isoscalar part is proportional to the total spin operator.
This means that the total spin is conserved in such magnetic transitions.
Isoscalar $M1$ transitions are forbidden from initial $S$ waves if distortion is neglected because of the orthogonality between initial and final states.
More generally, these transitions are forbidden if the initial $S$ component is an eigenstate of the spin operator,
\beq
S_z \Psi_{i0}^{JM+} = M \Psi_{i0}^{JM+}.
\eeqn{5.1}
An isoscalar transition from the $S$ wave, which is expected to provide the largest contribution to the $M1$ $S$-factor, is not possible without components with total spin $S = 0$ or total orbital momentum $\cL \geq 2$ in this $L = 0$ initial partial wave.

A transition from an $S$ wave is nevertheless also made possible by isospin 1 components in the wave functions.
Given the fact that the shapes of $D$ wave $S$-factors at small energies are affected by centrifugal barriers,
this suggests that such partial $S$-factors decrease when the energy tends to zero.
Even with a weak $T = 1$ component, transitions from the initial $S$ partial wave can dominate at the lowest energies.

These findings are illustrated by the $p + n + \alpha$ three-body model.
Both isoscalar and isovector $M1$ contributions from an initial $D$ wave are negligible in the astrophysical energy domain.
Isovector $M1$ transitions from an $S$ wave are much more important at small energies because of their rather flat behavior but they are restricted by their dependence on small isospin 1 components in the final wave function.
The smallness of the $M1$ $S$-factor seems therefore quite normal and agrees with some previous calculations \cite{NWS01,So22}.

Improvements of this three-body model might enhance the $M1$ $S$-factor.
A more elaborate initial scattering wave function than \Eq{3.5} should read
\beq
\Psi_i^{JM+} = \sum_\gamma [[Y_l(\Omega_r) \otimes Y_L(\Omega_R)]^{\cL} \otimes \chi_S]^{JM} \psi^J_{i\gamma}(r,R),
\eeqn{5.2}
where $\gamma = (lL)\cL S$ and $l + L$ is even.
This flexible three-body wave function corresponds to relaxing the frozen deuteron in the initial wave by allowing $p + n$ distortions.
Function $\psi^J_{i(00)01}(r,R)$ has the same asymptotic behavior as in \Eq{3.5}.
With \Eq{5.2}, \Eq{5.1} can be violated and isospin 1 components which are not present in this model can appear in the initial state for $l = 1$.
Such components will contribute to isovector transitions.

Transitions to the $^6$Li($1^+$) ground state from the $S$ wave involve initial components $\psi^1_{i(l0)lS}(r,R)$ with $l$ even.
As $S_f = 1$ is the largest spin component in the final state \cite{TDB07}, spin conservation suggests
that the main transitions will involve $S = 1$.
The main part of the transition should occur from $\psi^1_{i(00)01}(r,R)$ to $\psi^1_{f(00)01}(r,R)$.
But the overlap of these components is strongly constrained by the orthogonality condition
\beq
\sum_{lL\cL S} \int_0^\infty dR\, R^2 \int_0^\infty dr\,  r^2 \psi^1_{f(lL)\cL S}(r,R) \psi^1_{i(lL)\cL S}(r,R) = 0.
\eeqn{5.3}
Since terms with either $S = 0$, $l > 0$, or $L > 0$ in the sum are small, the $(00)01$ term of this is overlap should also be small.
It is thus not obvious that the size of the $M1$ $S$-factor would be increased much with this generalization.

The $D$ wave part might be enhanced by the introduction of a tensor force coupling the initial $S$ and $D$ waves
but this effect might be marginal on $S$-factors because of the $L = 2$ centrifugal barrier.

Microscopic cluster models allow extensions which include excitations of the $\alpha$ particle.
Distortion of the $T = 0$ part and small $T = 1$ components in these excitations could modify the $M1$ $S$-factor
but the effect of such complicated modifications cannot be predicted.

In the \textit{ab initio} calculation of \re{HHK22}, the $M1$ $S$-factor
is orders of magnitude larger than here and in Refs.~\cite{NWS01,So22}.
Its energy dependence is also very different.
These differences may come from a more realistic description of the $^6$Li nucleus by the no-core shell-model part of the wave function in \re{HHK22}.
It would be interesting to perform a similar calculation using the effective microscopic $M1$ operator presented in \Eq{2.5}.
An analysis of the various contributions to the $M1$ $S$-factor would provide information about the origin of its large value in that approach
and suggest possible improvements for other models.
\section{Conclusion}
\label{sec:conc}
The $M1$ contribution to cross sections of the $\alpha(d,\gamma)^6$Li radiative capture reaction
is subject to a controversy.
Here and in other models \cite{NWS01,So22}, it is found negligible
while it is important at low energies in an \textit{ab initio} calculation \cite{HHK22}.

To analyze this problem, we make use of an effective operator
exactly equivalent to the LWA form of the $M1$ operator
in radiative capture transition matrix elements \cite{Ba23}.
This operator can also simplify the computation of magnetic dipole moments by inverting \Eq{2.4}.
A general analysis of matrix elements of this operator for capture reactions shows
that isoscalar transitions from an initial $S$ wave are not allowed
if a rather general condition is satisfied [see \Eq{5.1}].
These transitions are forbidden by the orthogonality between the initial and final states.
Isovector radiative capture from the $S$ wave remains possible
but depends on the amount of isospin 1 mixing in the wave functions.

An evaluation in a three-body model involving proton, neutron and a structureless $\alpha$ is performed.
This three-body model is interesting as it offers flexibility
leading to simpler physical interpretations than elaborate models.
Its initial $^4$He + deuteron and final $^6$Li($1^+$) wave functions arise from different Hamiltonians and are properly orthogonalized.
The resulting $M1$ $S$-factor is very small as expected from the structure of the effective operator
and is thus negligible with respect to the $E2$ $S$-factor.
It is too weak to contribute to the capture process at the lowest energies where experiments are available.
This confirms results of the partly phenomenological six-body model of \re{NWS01} and the microscopic cluster model of \re{So22}.
Some aspects of the present three-body study which can limit its predictive power are discussed.

The large size difference of $M1$ capture between the \textit{ab initio} calculation of \re{HHK22} and other models remains unexplained.
It would be interesting to have detailed \textit{ab initio} results using the effective $M1$ operator presented in \Eq{2.5}.
They would explain the physical origin of the large $M1$ $S$-factor in that approach.
%
%
\appendix
\section{Reduced matrix elements in the three-body model}
\label{sec:A}
The radial part of the final $^6$Li wave function \rref{3.4} for $J_f = 1$  and $\gamma = (l_fL_f)\cL_fS_f$ reads
\beq
\psi^{J_f}_{f\gamma}(r,R) = \rho^{-5/2} \sum_{K} \chi_{\gamma K}(\rho) \Phi_K^{l_fL_f}(\alpha)
\eeqn{A1}
where $\rho$ and $\alpha$ are hyperspherical coordinates, the quantum number $K$ is the hypermomentum,
and the hyperradial function $\chi_{\gamma K}$ is the result of the Lagrange-mesh calculation.
The hyperangular part is defined as
\beq
\Phi_K^{l_fL_f}(\alpha) = N_K^{l_fL_f} (\cos \alpha)^{l_f} (\sin \alpha)^{L_f} P_{n}^{L_f+1/2,l_f+1/2} (\cos 2\alpha),
\eeqn{A2}
where $P_{n}^{L_f+1/2,l_f+1/2}$ is a Jacobi polynomial with $n$ given by $K = l_f+L_F+2n$
and $N_K^{l_fL_f}$ is a normalization factor \cite{DDB03}.
The hyperradius is given by $\rho = \sqrt{\dem r^2 + \qt R^2}$ and the hyperangle by $\alpha = \arctan (\sqrt{8/3} R/r)$.

When $J = J_f$, the initial wave function \rref{3.5} is orthogonalized to the final one.
The radial part of the initial $p + n + \alpha$ wave function for relative orbital momentum $L$ is modified as
\beq
\psi^{J}_{iL}(r,R) = (rR)^{-1} u_d(r) g^J_L(R) - \delta_{J_fJ} \delta_{L_fL} \delta_{l_f0} \psi^{J_f}_{f\gamma}(r,R)
\eol
\times \int_0^\infty dR\, R g^J_L(R) \int_0^\infty dr\, r u_d(r) \psi^{J_f}_{f\gamma}(r,R).
\eeqn{A3}

The reduced matrix element of the spin operator in the isoscalar part of $\cM^{M1}$ is given by
\beq
\la \Psi^{J_f}_{f\gamma}|| S || \Psi^{J}_{iL} \ra
= \delta _{S_f1} \delta _{l_f0} \delta_{\cL_fL} \delta_{L_fL} (-1)^{L+J_f}\left[6(2J+1) \right]^{1/2}
\left\{\begin{array}{ccc} J_f & J & 1 \\ 1 & 1 & L \end{array}\right\} I_O
\eeqn{A5}
where the overlap integral reads
\beq
I_O = \int_0^\infty dR\, R^2 \int_0^\infty dr\, r^2 \psi^{J_f}_{f\gamma}(r,R) \psi^{J}_{iL}(r,R).
\eeqn{A4}
The reduced matrix element of the spin scalar operator in the isovector part of $\cM^{M1}$ in Eqs.~\rref{3.10} and \rref{3.11} reads
\beq
\la \Psi^{J_f}_{f\gamma} || \qt \vR \times \vp + \dem \vq \times \vP ||\Psi^{J}_{iL} \ra & = & \delta_{S_f1} \delta_{l_f1}
(-1)^{\cL_f+J+L_f} \left[ 3(2L_f+1)(2J+1)(2L+1) \right]^{1/2}
\eol && \times
\left\{\begin{array}{ccc} J_f & \cL_f & 1 \\ L & J & 1 \end{array}\right\}
\left\{\begin{array}{ccc} 1 & 1 & 1 \\ \cL_f & L_f & L \end{array}\right\}
\left(\begin{array}{ccc} L_f & L & 1 \\ 0 & 0 & 0 \end{array}\right) I
\eeqn{A6}
with the radial integral
\beq
 I & = & \int dR\, R^2 \int dr\, r^2 \psi^{J_f}_{f\gamma}(r,R)
\eol && \times \left[ -\frac{4}{3} R\frac{d}{dr} + \frac{1}{2} r \frac{d}{dR}
- r \frac{L_f(L_f+1)-L(L+1)-2}{4R} \right] \psi^{J}_{iL}(r,R).
\eeqn{A7}
The reduced matrix element of the spin vector operator in the isovector part of $\cM^{M1}$ is
\beq
\la \Psi^{J_f}_{f\gamma} || \vs_p - \vs_n || \Psi^{J}_{iL} \ra
= \delta _{S_f0} \delta _{l_f0} \delta_{\cL_fL} \delta_{L_fL} (-1)^{J_f+J}\, 2\left( \frac{2J+1}{3(2L+1)} \right)^{1/2} I_O.
\eeqn{A8}

\end{document}